\documentclass[aps,pra,showpacs,superscriptaddress,groupedaddress]{revtex4}  

\usepackage{graphicx}   
\usepackage{bm}             
\usepackage{amsmath}   
\usepackage{amssymb}   
\usepackage{epsfig}
\usepackage{color}
\newcommand{\blue}[1]{{\color{black} #1}}

\begin{document}


\title{Dissociative electron attachment cross sections for \blue{ro-}vibrationally excited NO molecule and N$^-$ anion formation}

\author{V.~Laporta}
\email{vincenzo.laporta@istp.cnr.it}
\affiliation{Istituto per la Scienza e Tecnologia dei Plasmi, CNR, Bari, Italy}

\author{I.F.~Schneider}
\affiliation{Laboratoire Ondes et Milieux Complexes, CNRS--Universit{\'{e}} Le Havre Normandie, Le Havre, France}
\affiliation{Laboratoire Aim{\'{e}} Cotton, CNRS--Universit{\'{e}} Paris-Saclay, ENS Cachan, Orsay, France}

\author{J.~Tennyson}
\affiliation{Department of Physics and Astronomy, University College London, London, London
WC1E 6BT, UK}

\begin{abstract}
Motivated by the huge need of data for non-equilibrium plasma modeling, a theoretical investigation of dissociative electron attachment to the NO molecule is performed. The calculations presented here are based on the Local-Complex-Potential approach, taking into account five NO$^-$ resonances. Three specific channels of the process are studied, including the production of excited nitrogen atoms $\mathrm{N}(^2\mathrm{D})$ and of its anions  N$^-$. Interpretation of the existing experimental data and their comparison with our theoretical result are given. A full set of \blue{ro-}vibrationally-resolved cross sections and the corresponding rate coefficients are reported. In particular, a relatively notably large cross section of N$^-$ ion formation at low energy of the incident electron and for vibrationally excited NO target is predicted. \blue{Finally, molecular rotation effects are  discussed.}
\end{abstract}

\pacs{xxxxx}

\maketitle


In our recent paper~\cite{10.1088/1361-6595/ab86d8} we produced theoretical cross sections for the dissociative excitation of the nitric oxide (NO) molecule by electron impact. In order to supply further data for the non-equilibrium plasma modeling, and based on the same molecular data, in this letter we extend the calculations to its dissociative electron attachment (DA).

Although the NO molecule is only a minor constituent of Earth's upper atmosphere, it plays a major role in the infrared aurora borealis emissions~\cite{doi:10.1029/2003GL019151} due to radiation coming from its vibrationally excited states, and in the energy transfers in the boundary layer of the plasma created in front of the spacecrafts entering in the planetary ionospheres~\cite{:/content/aip/journal/pop/20/7/10.1063/1.4810787, doi:10.1063/1.4904817, doi:10.1063/1.4900508, laporta:104319, Laporta201644}. Moreover, among the nitrogen oxide compounds $\mathrm{N}_x\mathrm{O}_y$, the NO molecule and its radicals have the greater impact on the environment and on the pollution caused by human activities~\cite{Kreuzer45, Spicer:1977aa}, being also very important in many industrial technologies~\cite{doi:10.1080/0144235X.2016.1179002, Campbell_2012, Motapon-PSST-2006, doi:10.1029/2002JA009458} and in the optimization of combustion processes of the fossil fuels~\cite{SLAVCHOV2020117218, doi:10.1080/00102206908952211}.

Many theoretical studies~\cite{vivie:3028, 0963-0252-21-5-055018, PhysRevA.71.052714, PhysRevA.69.062711} and experimental investigations~\cite{PhysRev.172.125, 0953-4075_38_5_011, rapp:1480, PhysRevLett.90.203201, Josic2001318, Krishnakumar_1988, PhysRevA.10.1633} on the electron-NO reactions are available and recently reviewed by Song {\it et~al.}~\cite{doi:10.1063/1.5114722} and by McConkey {\it et~al.}~\cite{McConkey20081}. Specifically, concerning the DA process, the theoretical studies are limited to very low scattering energy and to the few lowest vibrational states corresponding to the ground electronic state of NO, and only few low-lying resonant states of NO$^-$~\cite{PhysRevA.71.052714} are taken into account. The only absolute experimental DA cross section measurements were performed by Rapp and Briglia ~\cite{rapp:1480} who observed O$^-$ production from the $v=0$ vibrational state of NO. Experiments have detected both excited nitrogen atoms and N$^-$ production from DA~\cite{doi:10.1063/1.459882, PhysRevLett.74.5017, C0CP01067G, DENIFL1998105} but these observations are not yet understood. Moreover, accurate collisional-radiative models of plasmas, which are able to investigate internal energy exchanges, require a state-to-state approach, which consists of cross sections resolved over the internal degrees of freedom for the consituent atoms and molecules~\cite{doi:10.1063/1.4904817, doi:10.1063/1.4900508}.

In this letter we present new calculations of \blue{ro-}vibrational state-resolved cross sections and  the corresponding rate coefficients for the following three resonant DA processes:
\begin{align}
e(\epsilon) + \mathrm{NO}(\mathrm{X}\,^2\Pi; v\blue{, j}) &\to \mathrm{NO}^- \to \mathrm{N}(^4\mathrm{S}) + \mathrm{O}^-(^2\mathrm{P})\,,&\mathrm{ (DA1)} \label{eq:DAprocess1}\\
e(\epsilon) + \mathrm{NO}(\mathrm{X}\,^2\Pi; v\blue{, j}) &\to \mathrm{NO}^- \to \mathrm{N}^-(^3\mathrm{P}) + \mathrm{O}(^3\mathrm{P})\,,& \mathrm{(DA2)} \label{eq:DAprocess2}\\
e(\epsilon) + \mathrm{NO}(\mathrm{X}\,^2\Pi; v\blue{, j}) &\to \mathrm{NO}^- \to \mathrm{N}(^2\mathrm{D}) + \mathrm{O}^-(^2\mathrm{P})\,,& \mathrm{(DA3)} \label{eq:DAprocess3}
\end{align}
\blue{where $v$ and $j$ represent the vibrational and rotational quantum number of the NO molecule. In the following, where not explicit specified, we assume $j=0$; rotational effects are discussed in the last part of the letter.} Since the excited electronic states of nitrogen atoms as well as its negative ion are unstable they undergo decays toward the ground state $\mathrm{N}(^4\mathrm{S})$: in the plasma, the reaction DA2 will be followed by detachment process by electron emission for N$^-$~\cite{Mazeau_1978} and DA3 channel will be followed by radiative decay to N$(^2\mathrm{D})$.

As specified in~\cite{10.1088/1361-6595/ab86d8}, in order to cover a large energy range for the incident electron, we take into account five resonant electronic states of NO$^-$:  three low-lying  -- \textit{i.e.} of $^3\Sigma^-$, $^1\Sigma^+$ and $^1\Delta$ symmetries -- and two higher -- of $^3\Pi$ and $^1\Pi$ symmetries -- close to the NO dissociation threshold. In the following, we will refer to these resonances by $r=1,\ldots,5$. The correspondence between each resonance $r$ and a specific channel $i=1,2,3$, associated to an asymptotic (dissociation) limit, is reported in Table~\ref{tab:NO-_limits}, which also gives the electron affinity for each process.

\begin{table}
\caption{Asymptotic limit positions of the NO$^-$ resonant  states considered in the calculations for the three DA channels (\ref{eq:DAprocess1})-(\ref{eq:DAprocess3}). The energy is expressed with respect to the asymptotic limit of the ground electronic state of NO.  Experimental energy values, from Mazeau \textit{et al.}~\cite{Mazeau_1978} and NIST database~\citep{NIST_ASD}, are given in brackets for comparison. The last column reports the correlation with the resonances for each channel. \label{tab:NO-_limits}}
\centering
\begin{tabular}{cccl}
\hline
~~Channel~~  & Limit & ~~~~Energy (eV)~~~~ & Symmetries\\
\hline
DA1 & $\mathrm{N}(^4\mathrm{S}) + \mathrm{O}^-(^2\mathrm{P})$ & --1.460  (--1.46) &  $^3\Sigma^-$\\
DA2 & $\mathrm{N}^-(^3\mathrm{P}) + \mathrm{O}(^3\mathrm{P})$ & +0.068  (+0.07) &  $^1\Delta$, $^1\Sigma^+$\\
DA3 & $\mathrm{N}(^2\mathrm{D}) + \mathrm{O}^-(^2\mathrm{P})$ & +0.910 (+0.91) &  $^3\Pi$, $^1\Pi$\\
\hline
\end{tabular}
\end{table}



According to the theoretical formalism of the Local-Complex-Potential (LCP) -- previously used to study the NO dissociative excitation process in~\cite{10.1088/1361-6595/ab86d8} and the DA of O$_2$ ~\cite{PhysRevA.91.012701} -- the DA cross section for an NO molecule initially on its vibrational level $v$  colliding with an electron of incident energy $\epsilon$
and following channel $i$ is given by~\cite{PhysRevA.20.194}:
\begin{equation}
\sigma^i_v(\epsilon) = \sum_{r} \frac{2S_r+1}{(2S+1)\,2} \frac{g_r}{g\,2} 
2\pi^2\,\frac{K}{\mu}\,\frac{m_e}{k}\,\lim_{R\to\infty}\left|\xi^r_v(R)\right|^2\,, \hspace{1cm}i=1,2,3 \,,  \label{eq:DeA_xsec}
\end{equation}
where $2S_r+1$ and $2S+1$ are the spin-multiplicities of the resonant electronic state of the anion and of the electronic state of the neutral target respectively, $g_r$ and $g$ represent the corresponding degeneracy factors, $K$ is the asymptotic momentum of the final dissociating fragments, $k$ is the incoming electron momentum, $\mu$ is the reduced mass of NO nuclei, $m_e$ is the electron mass and $\xi^r_v$ is the resonant wave function solution of the nuclear equation with total energy $E=\epsilon_v+\epsilon$:
\begin{equation}
\left[ -\frac{\hslash^2}{2\mu}\frac{d^2}{dR^2} + V_r^-(R)  - \frac{i}{2}\Gamma_r(R) - E \right]\xi^r_v(R) = -\mathcal{V}_r(R)\,\chi_v(R)\,, \hspace{1cm}r=1,\ldots,5 \,. \label{eq:res_wf}
\end{equation}
In Eq.~(\ref{eq:res_wf}), $V_r^-$ and $\Gamma_r$ represent the potential energies and the autoionization widths respectively for the five resonant NO$^-$ states included in the calculation, $\mathcal{V}_r$ the discrete-to-continuum coupling definied in~\cite{10.1088/1361-6595/ab86d8} and $\chi_v$ is the wave function of the initial vibrational state of NO with energy $\epsilon_v$:
\begin{equation}
\left[ -\frac{\hslash^2}{2\mu}\frac{d^2}{dR^2} + V_0(R) \right]\chi_v(R) = \epsilon_v\,\chi_v(R)\,,  \label{eq:NO_wf}
\end{equation}
where $V_0$ is the potential energy of the ground electronic state of NO$(\mathrm{X}\,^2\Pi)$. Full details concerning the LCP theoretical model are reported in~\cite{10.1088/1361-6595/ab86d8} and in the seminal papers~\cite{0034-4885-31-2-302, Domcke199197, PhysRevA.20.194}.

\begin{figure}
\centering
\includegraphics[scale=.39]{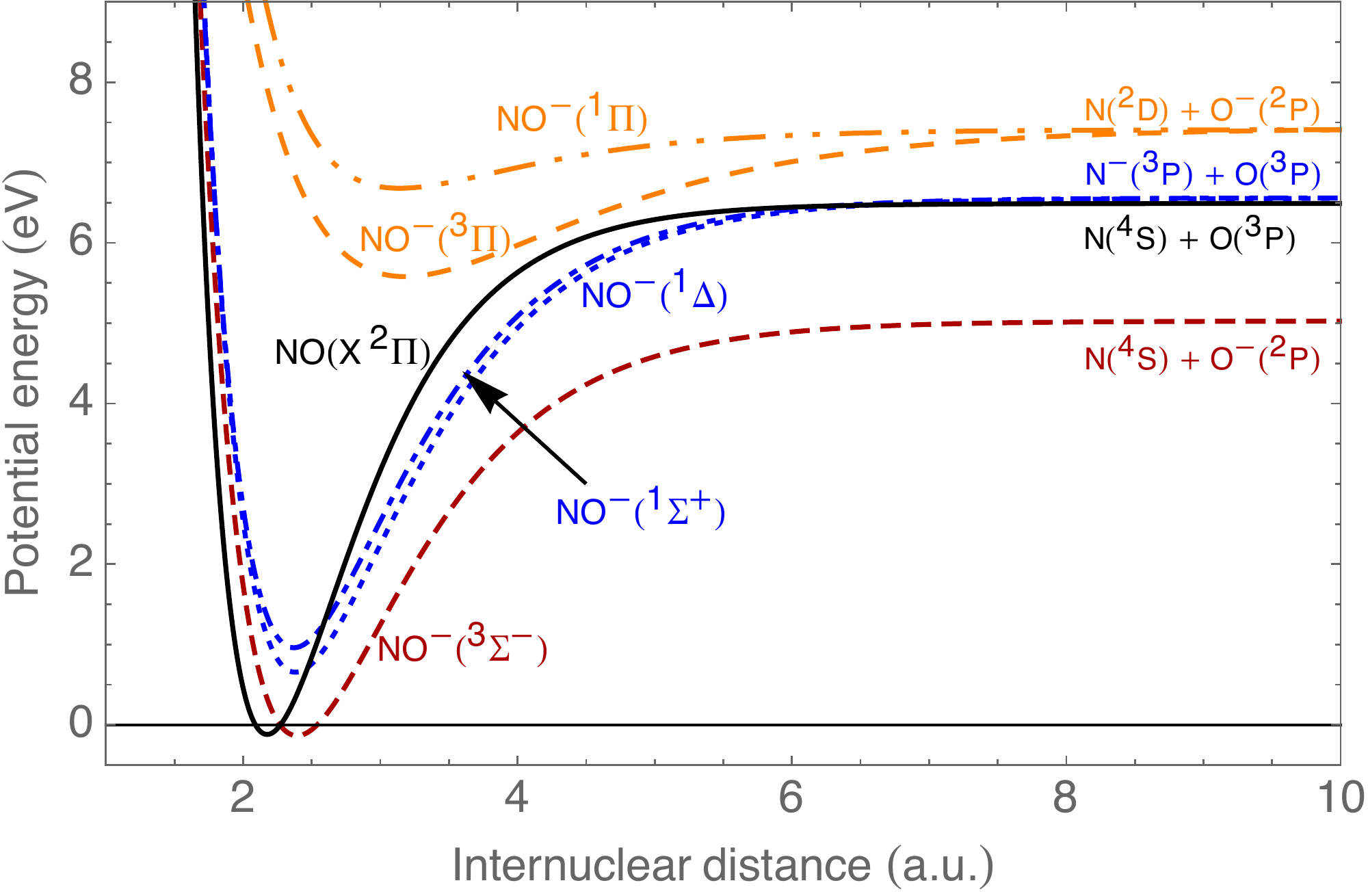} \hspace{.5cm} \includegraphics[scale=.4]{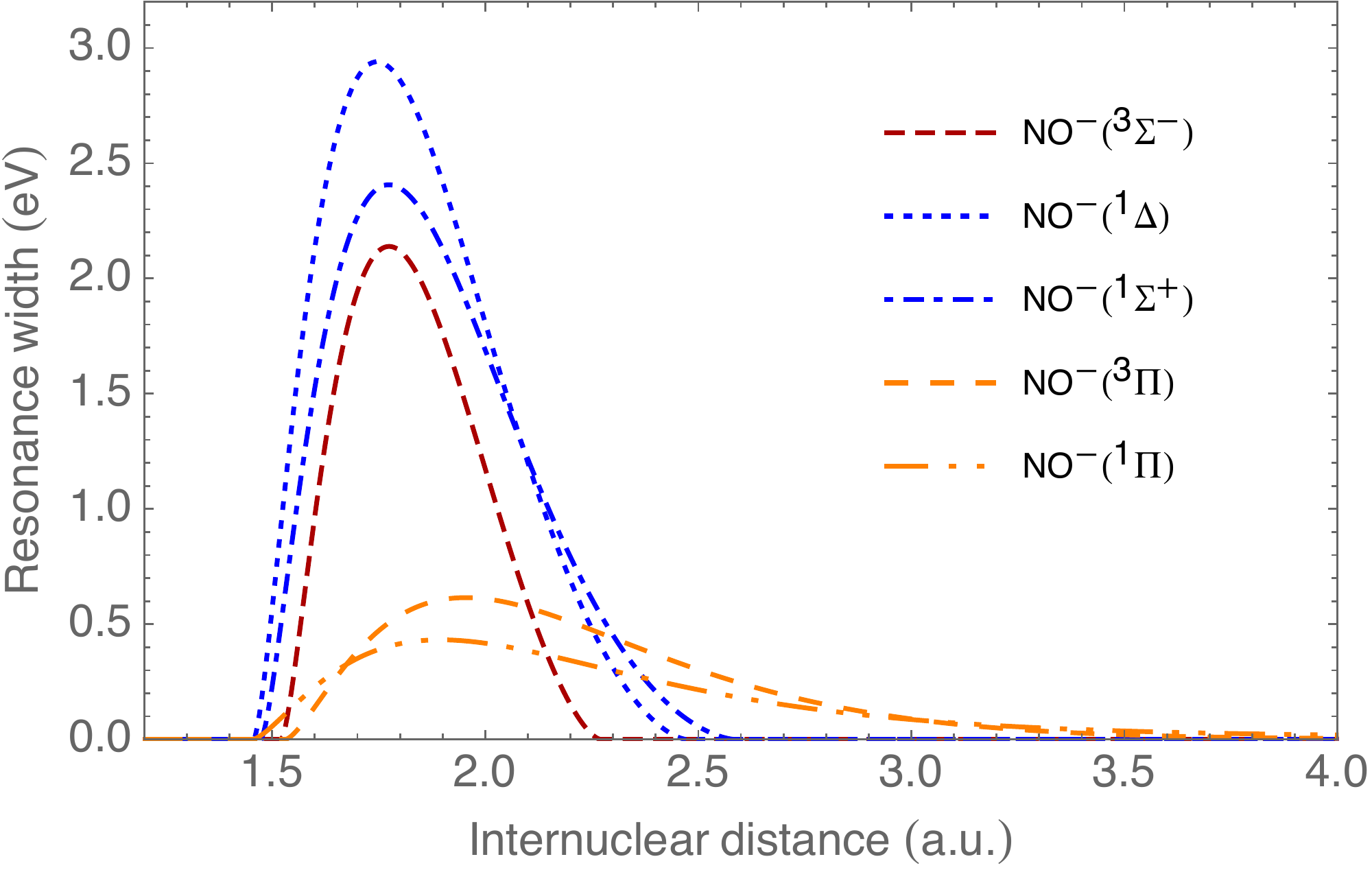} 
\caption{Molecular data as determined in~\cite{10.1088/1361-6595/ab86d8}. Left: Potential energy curves for the ground electronic state of NO molecule (solid line) and for the five NO$^-$ resonances (broken lines) included in the calculations, \blue{for $j=0$};  Right: The corresponding widths of the resonances. \label{fig:NOpot}}
\end{figure}

\begin{figure}
\centering
\includegraphics[scale=.28]{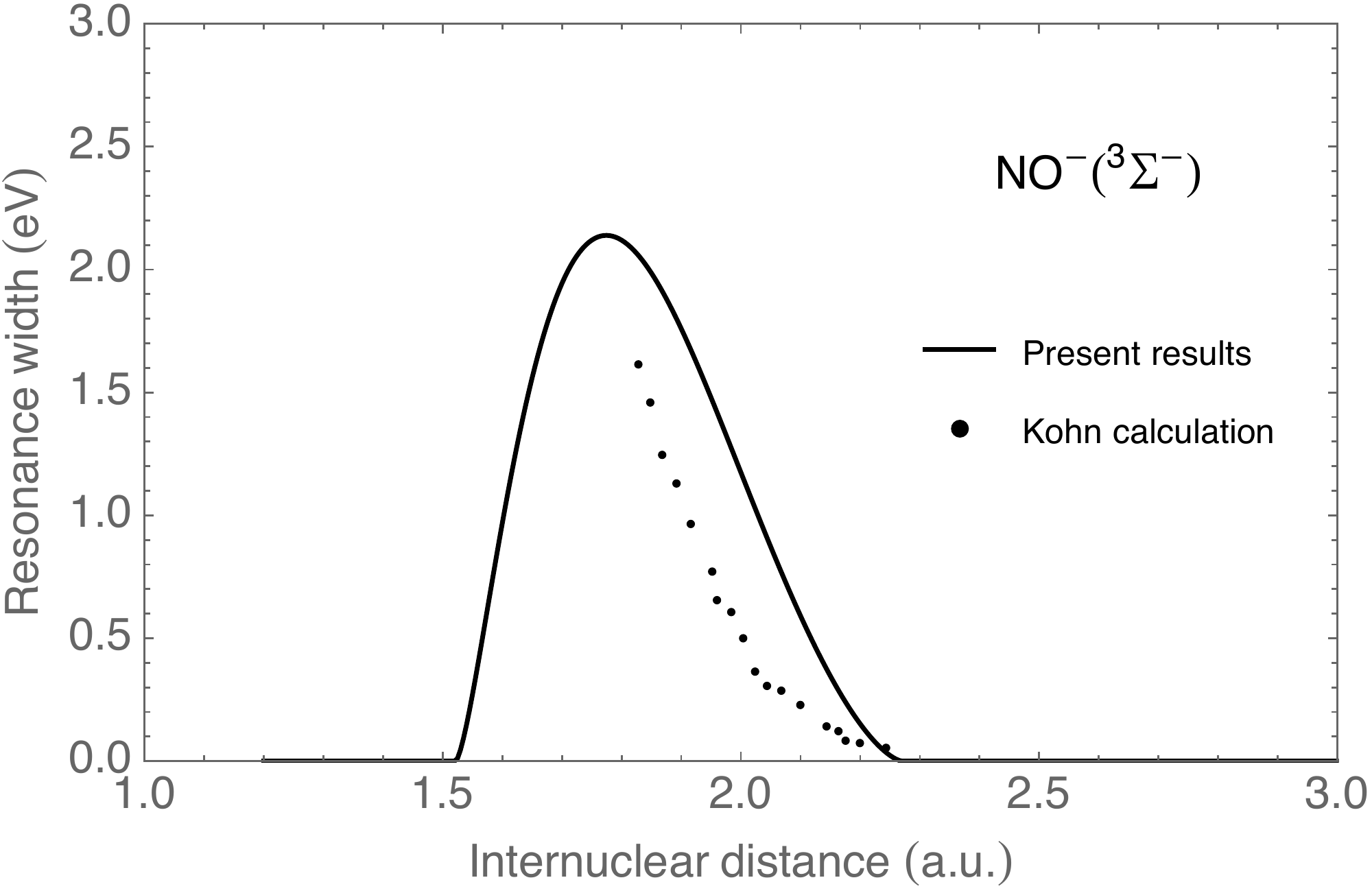} \includegraphics[scale=.28]{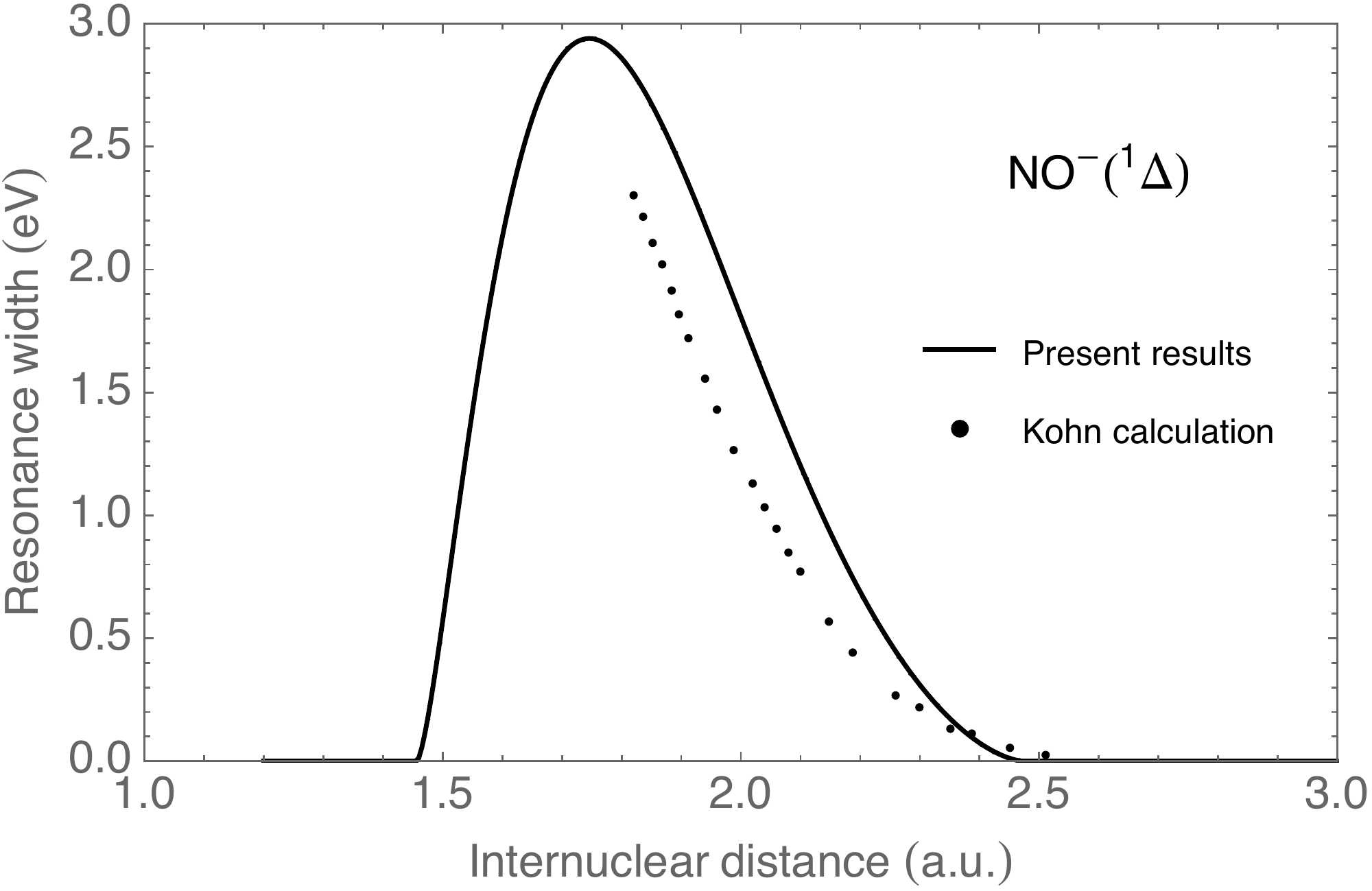} \includegraphics[scale=.28]{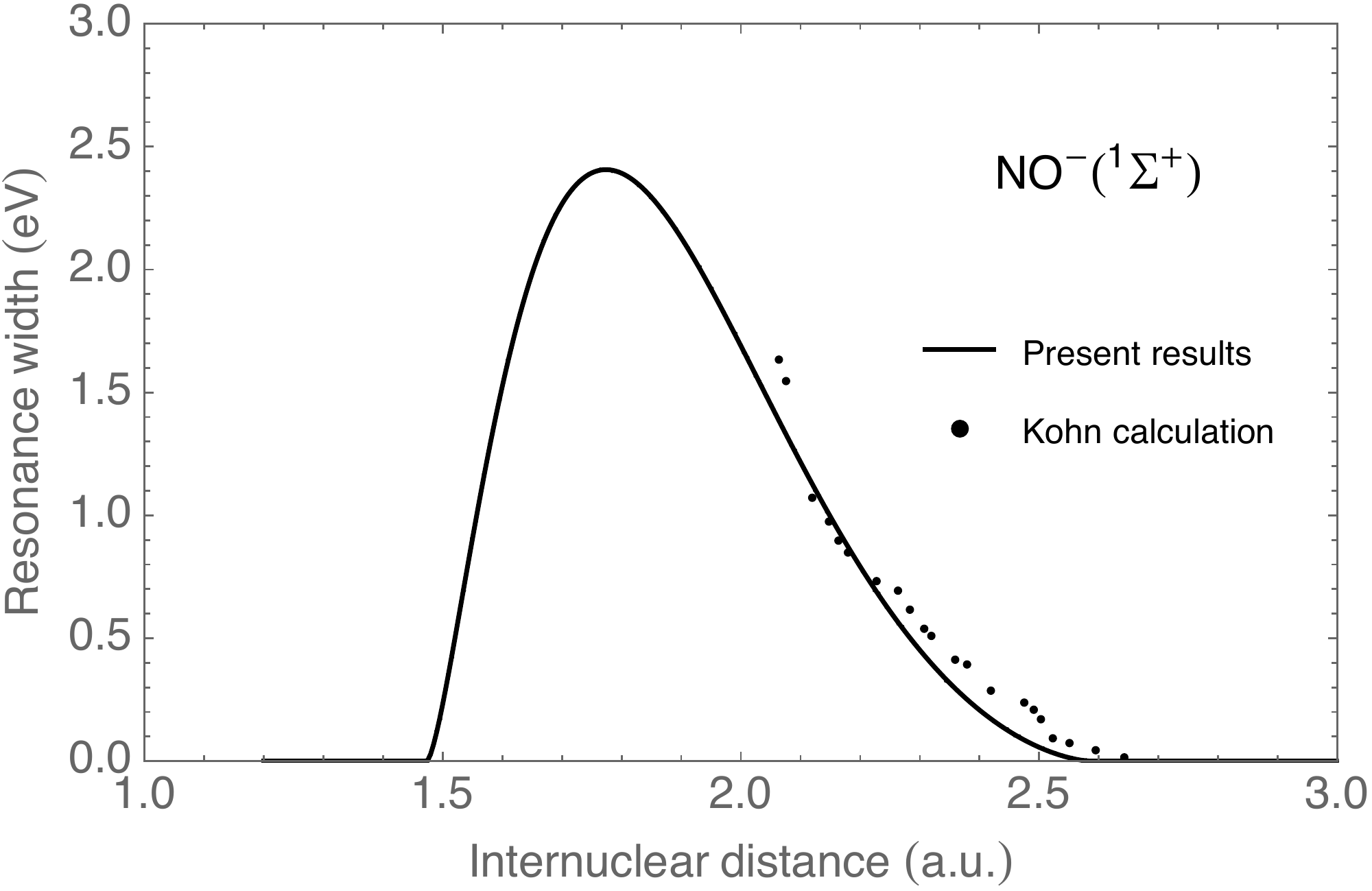} 
\caption{Autoionization widths of the low-lying NO$^-$ resonances: Comparison of the values used in the present calculation with those coming from the Kohn computations~\cite{PhysRevA.69.062711}.  \label{fig:gamma_cfr_rmatrix}}
\end{figure}

Figure~\ref{fig:NOpot} reports the complete set of  potential energy curves for NO in its ground electronic state, and the five NO$^-$ resonances and their corresponding autoionization widths. A discussion on molecular data determination as well as all input parameters entering in the Eqs.~(\ref{eq:DeA_xsec})-(\ref{eq:NO_wf}) can be found in~\cite{10.1088/1361-6595/ab86d8}.  In order to validate the resonance widths used in the present work and in~\cite{10.1088/1361-6595/ab86d8}, comparison with \textit{ab initio} Kohn variation principle results available in the paper~\cite{PhysRevA.69.062711} are shown in figure~\ref{fig:gamma_cfr_rmatrix}. Table~\ref{tab:NOviblev} contains the list of the vibrational levels $v$ supported by the NO molecule \blue{in its ground rotational state, {\it i.e.} $j=0$}.

\begin{table}
\centering
\begin{tabular}{cccccc}
\hline
~~~$v$~~~ & $~~\epsilon_{v}$~(eV)~~ & ~~~$v$~~~ & $~~\epsilon_{v}$~(eV)~~& ~~~$v$~~~ & $~~\epsilon_{v}$~(eV)~~\\
\hline
  0  &     0.000  &   18  &     3.581  &   36  &     5.745 \\ 
  1  &     0.236  &   19  &     3.739  &   37  &     5.823 \\ 
  2  &     0.468  &   20  &     3.892  &   38  &     5.897 \\ 
  3  &     0.695  &   21  &     4.040  &   39  &     5.967 \\ 
  4  &     0.918  &   22  &     4.185  &   40  &     6.033 \\ 
  5  &     1.137  &   23  &     4.324  &   41  &     6.094 \\ 
  6  &     1.351  &   24  &     4.460  &   42  &     6.150 \\ 
  7  &     1.561  &   25  &     4.591  &   43  &     6.203 \\ 
  8  &     1.767  &   26  &     4.718  &   44  &     6.251 \\ 
  9  &     1.968  &   27  &     4.840  &   45  &     6.294 \\ 
10  &     2.164  &   28  &     4.958  &   46  &     6.333 \\ 
11  &     2.357  &   29  &     5.072  &   47  &     6.368 \\ 
12  &     2.545  &   30  &     5.181  &   48  &     6.399 \\ 
13  &     2.729  &   31  &     5.286  &   49  &     6.425 \\ 
14  &     2.908  &   32  &     5.386  &   50  &     6.446 \\ 
15  &     3.083  &   33  &     5.483  &   51  &     6.464 \\ 
16  &     3.253  &   34  &     5.574  &   52  &     6.477 \\ 
17  &     3.419  &   35  &     5.662  &   53  &     6.485 \\ 
\hline
\end{tabular}
\caption{Energies of the vibrational levels of the electronic ground state of the NO molecule for \blue{$j=0$}. The dissociation energy for the ground vibrational level is $D_0=6.490$~eV \label{tab:NOviblev}}
\end{table}

\begin{figure}
\centering
\includegraphics[scale=.4]{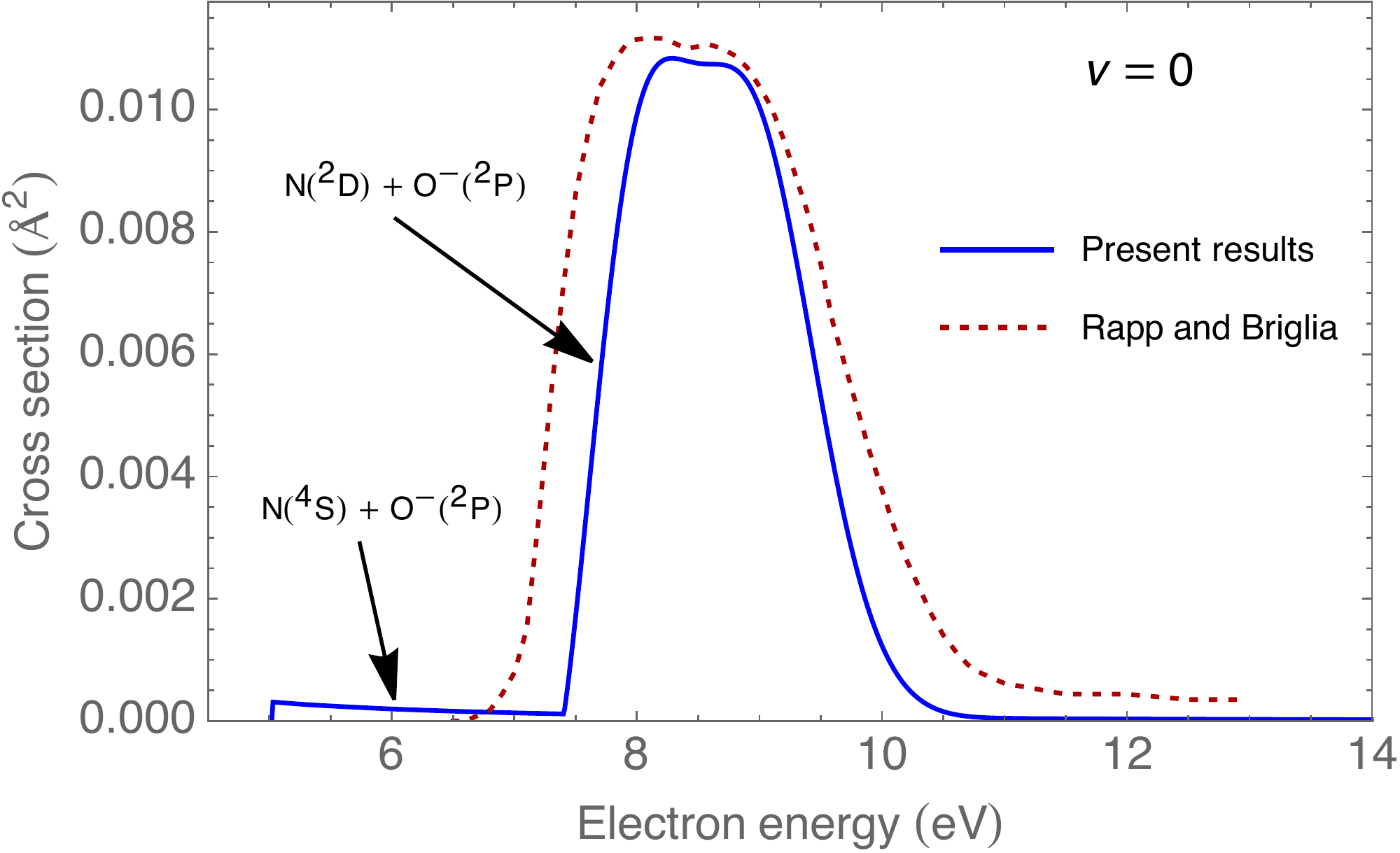} 
\caption{$\mathrm{O}^-$ production from DA1 and DA3 processes for $v=0$: Comparison of the present calculations \blue{ for $j=0$} with the experimental data of Rapp and Briglia~\cite{rapp:1480}. \label{fig:DAcfr_briglia}}
\end{figure}
\begin{figure}
\centering
\begin{tabular}{cc}
\includegraphics[scale=.28]{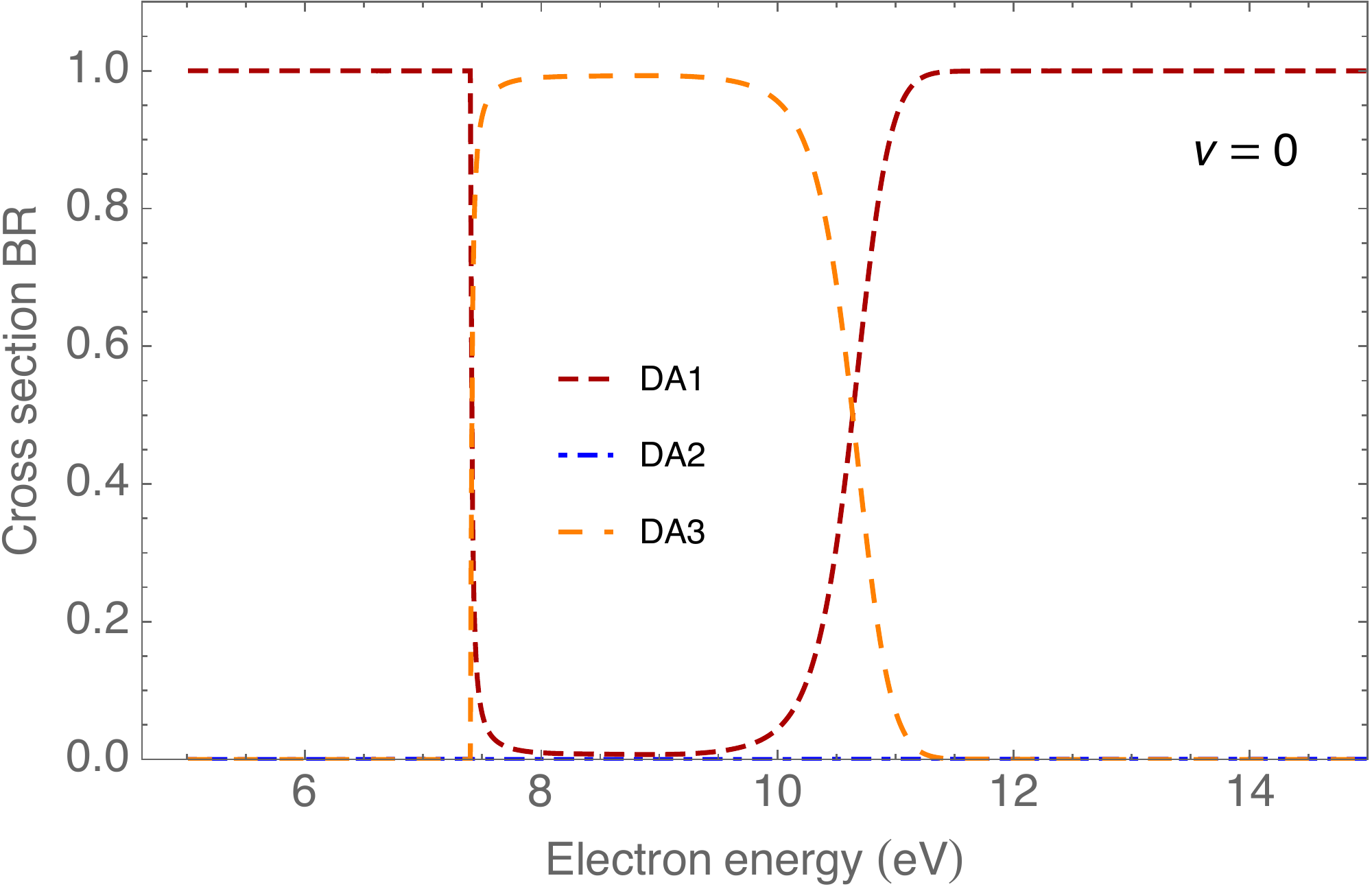} & \includegraphics[scale=.28]{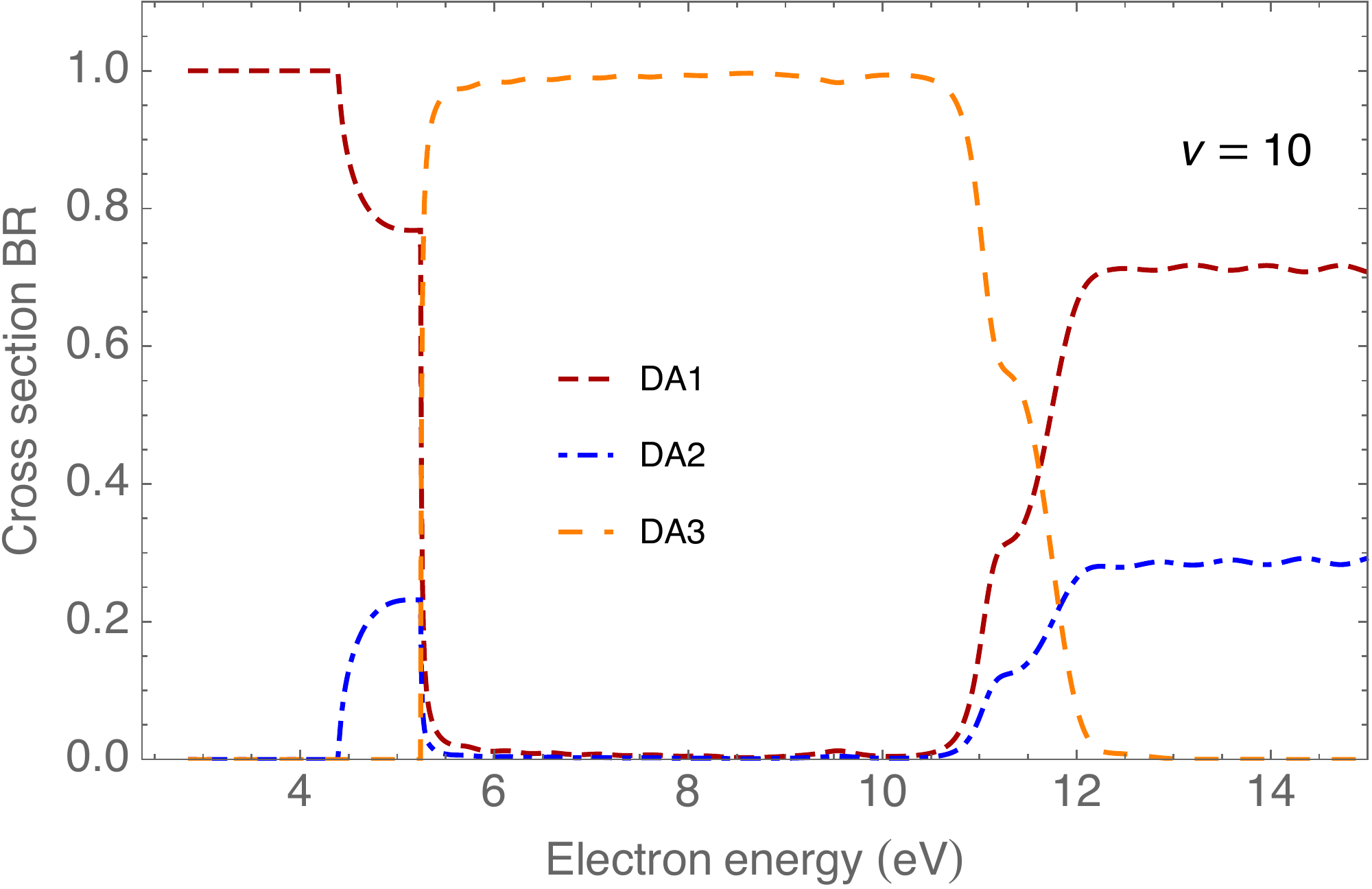} \\
\includegraphics[scale=.28]{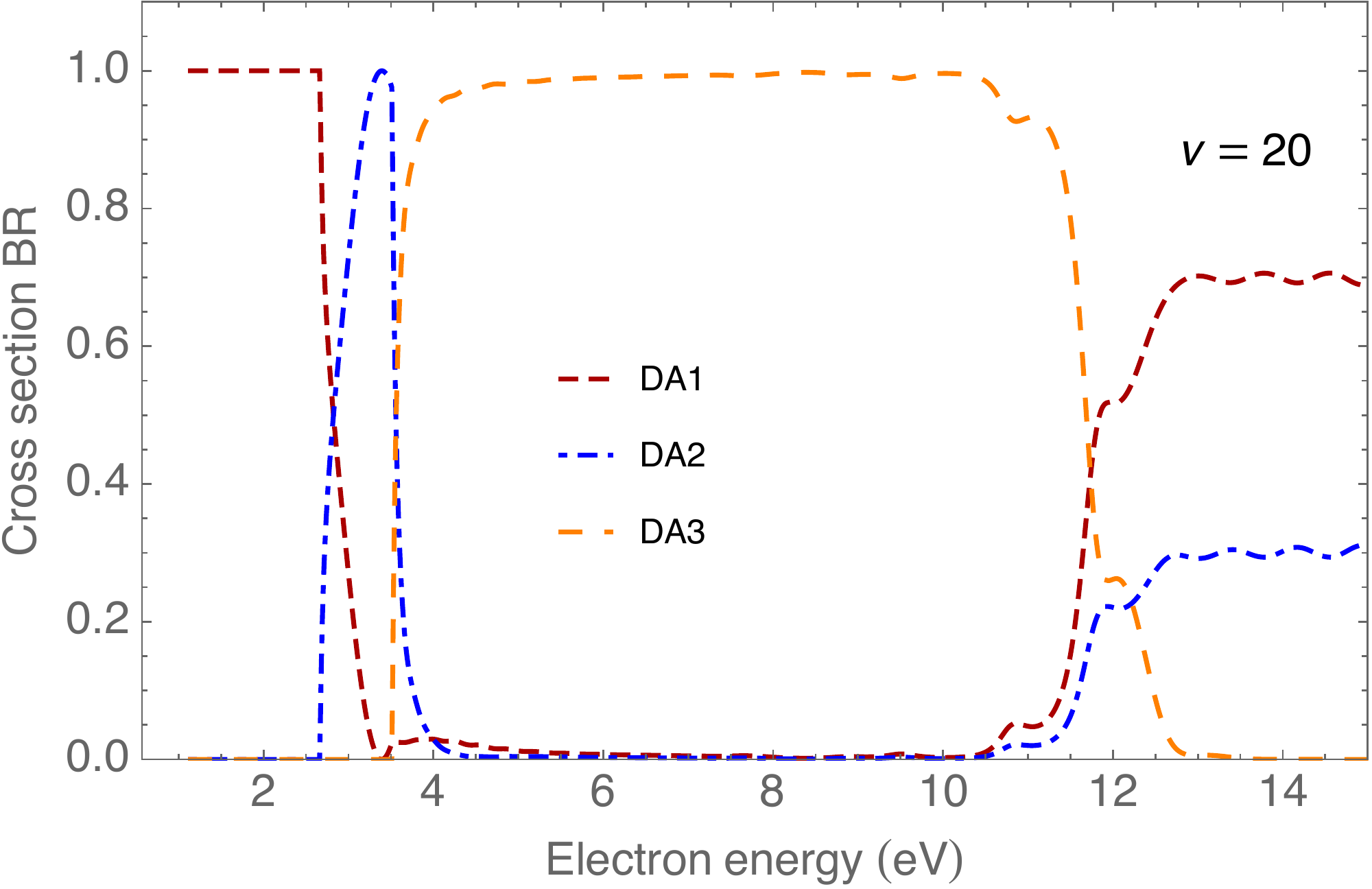} & \includegraphics[scale=.28]{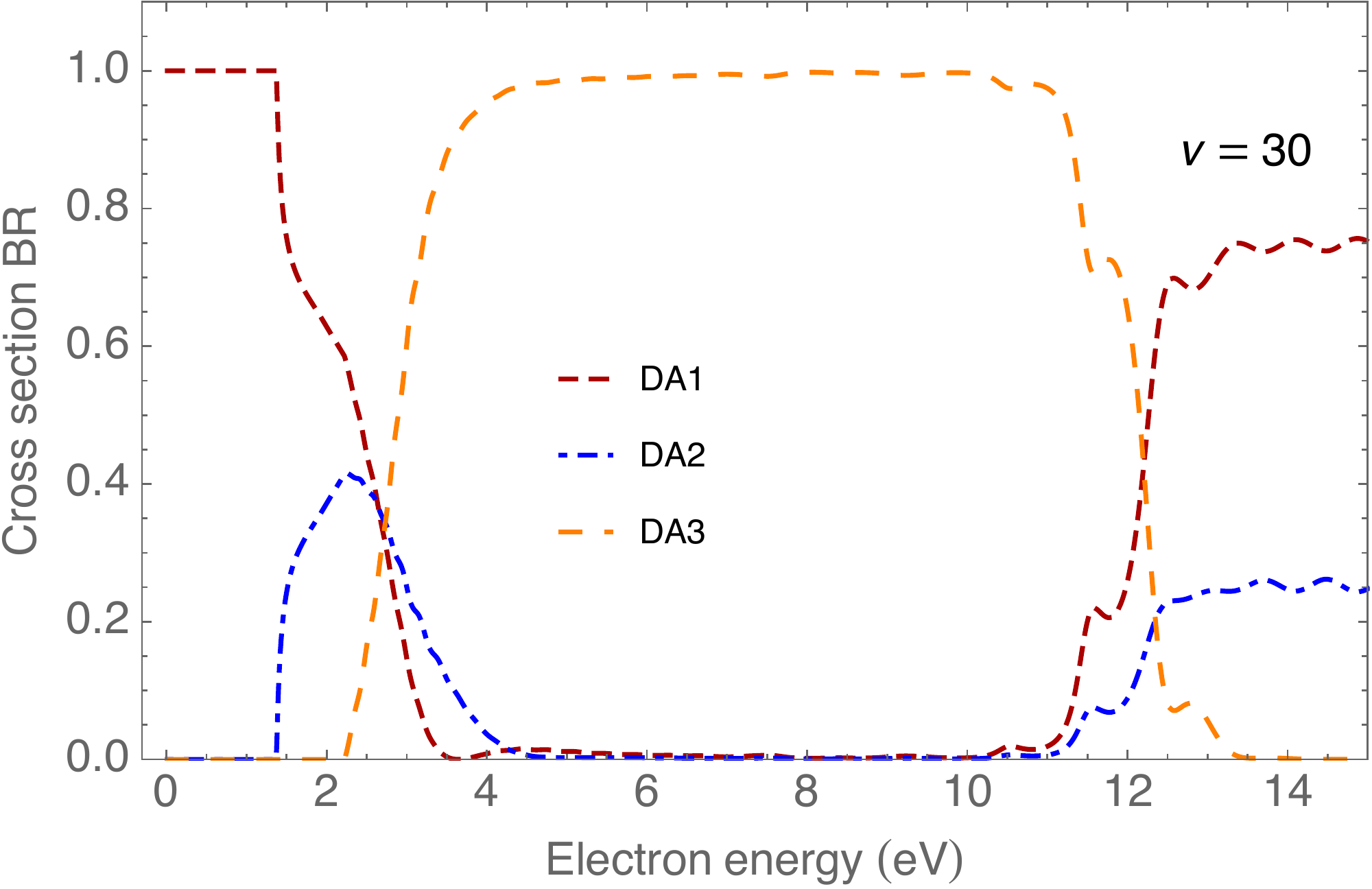} 
\end{tabular}
\caption{Cross section branching ratios for the three DA channels in (\ref{eq:DAprocess1}-\ref{eq:DAprocess3}) for $v=0$, $v=10$, $v=20$ and $v=30$ \blue{, for $j=0$}.  \label{fig:xsecDA_BR}}
\end{figure}
\begin{figure}
\centering
\begin{tabular}{ccc}
\includegraphics[scale=.28]{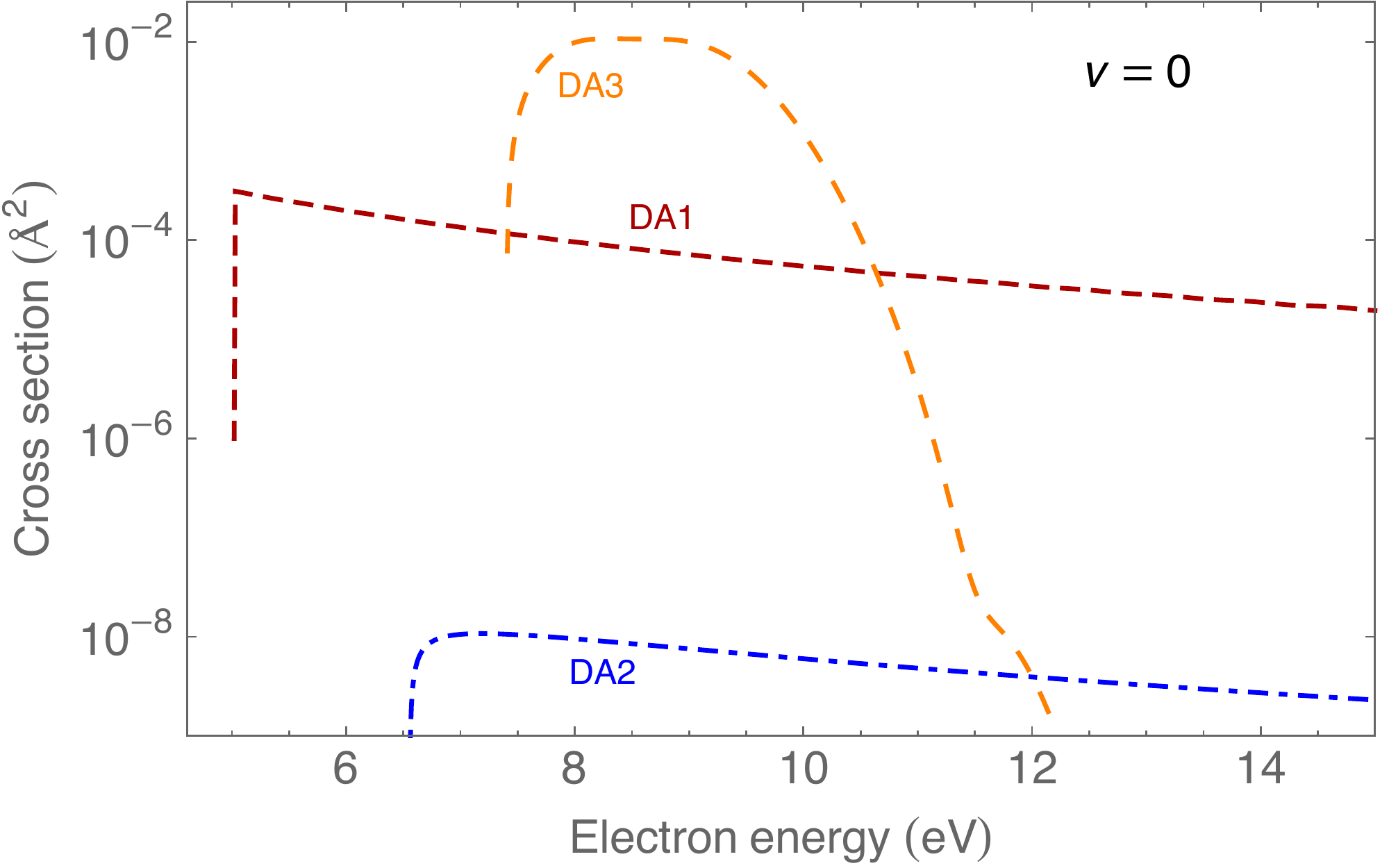} & \includegraphics[scale=.28]{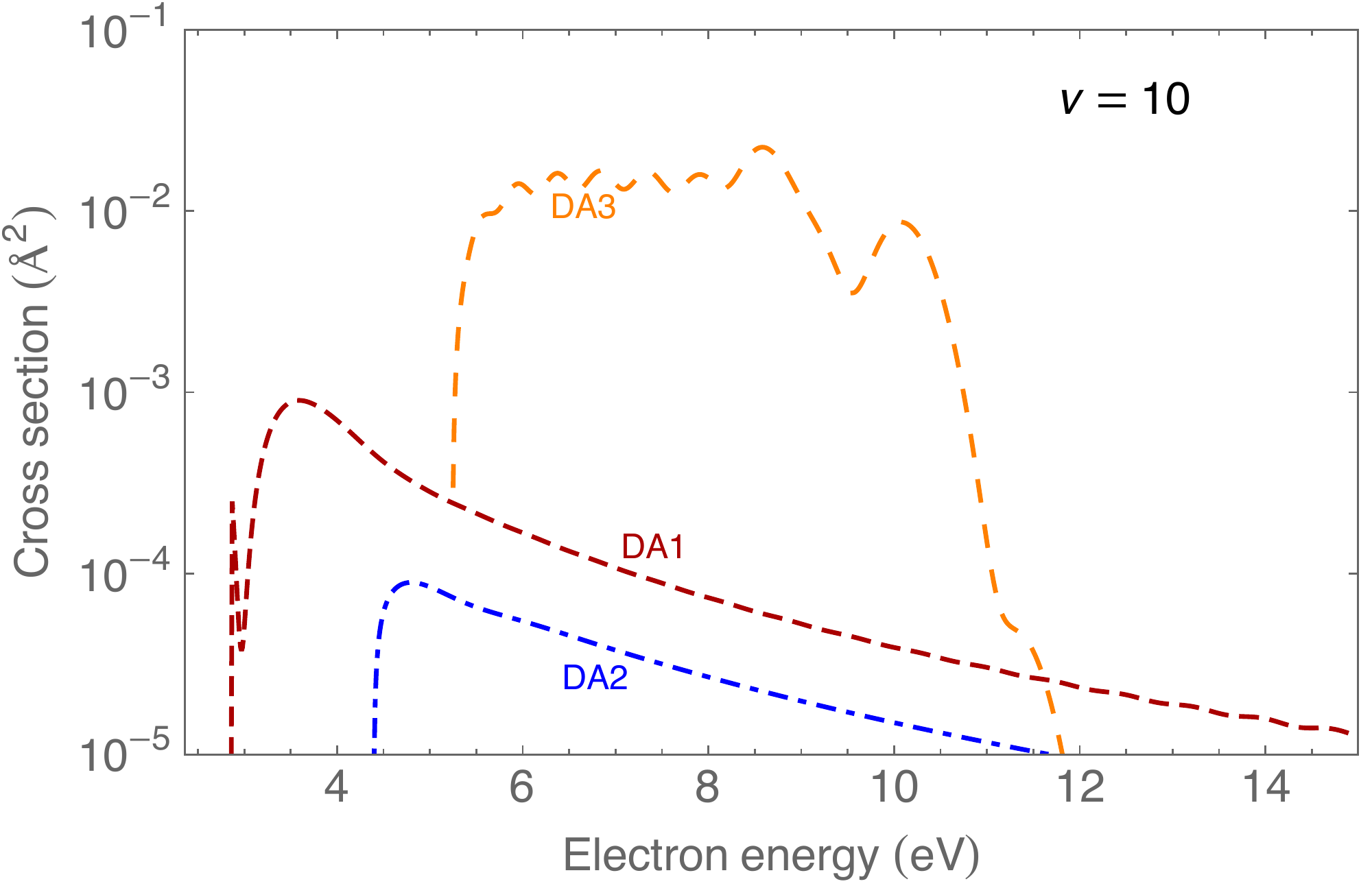} & 
\includegraphics[scale=.28]{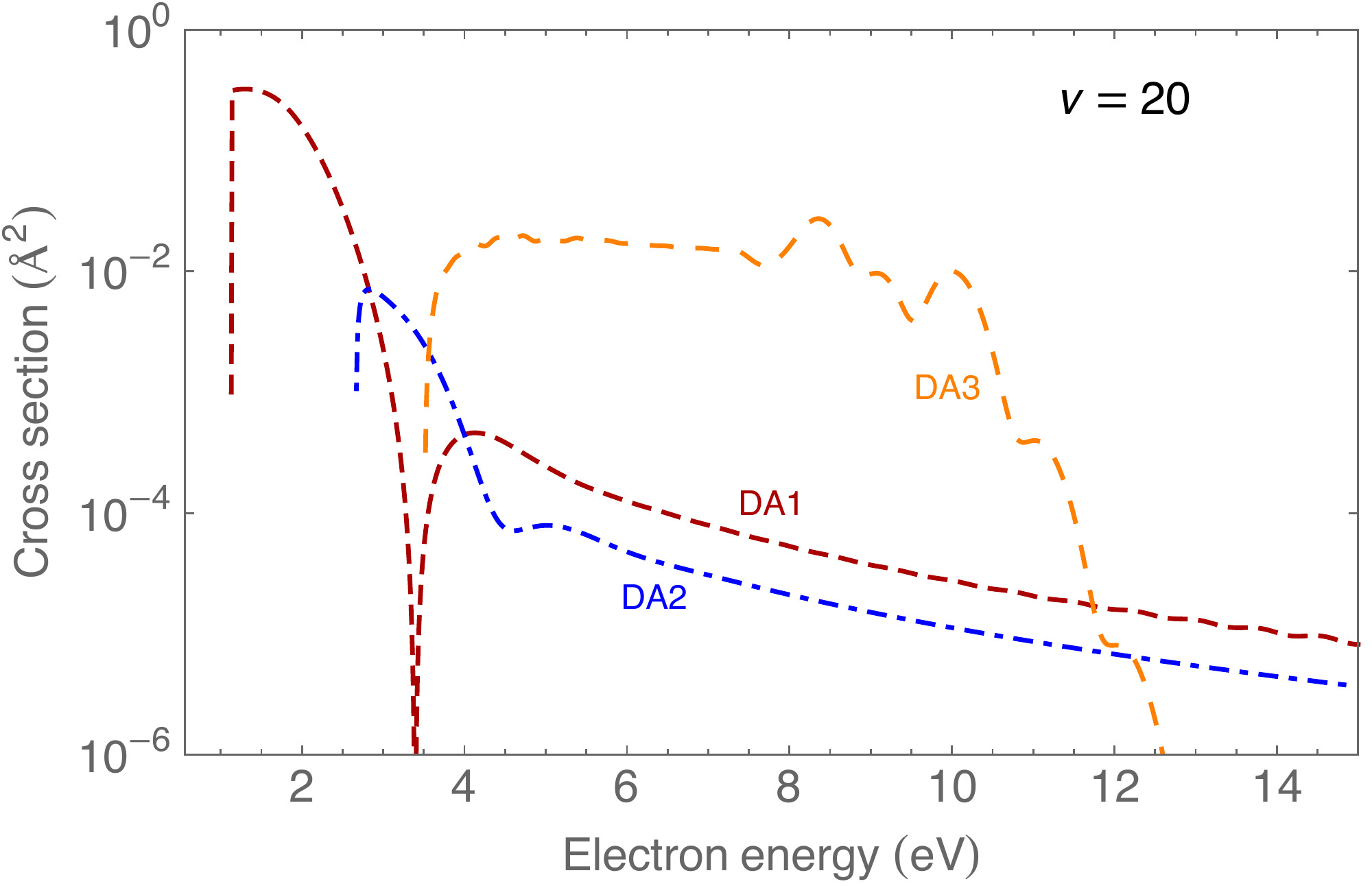}\\
\includegraphics[scale=.28]{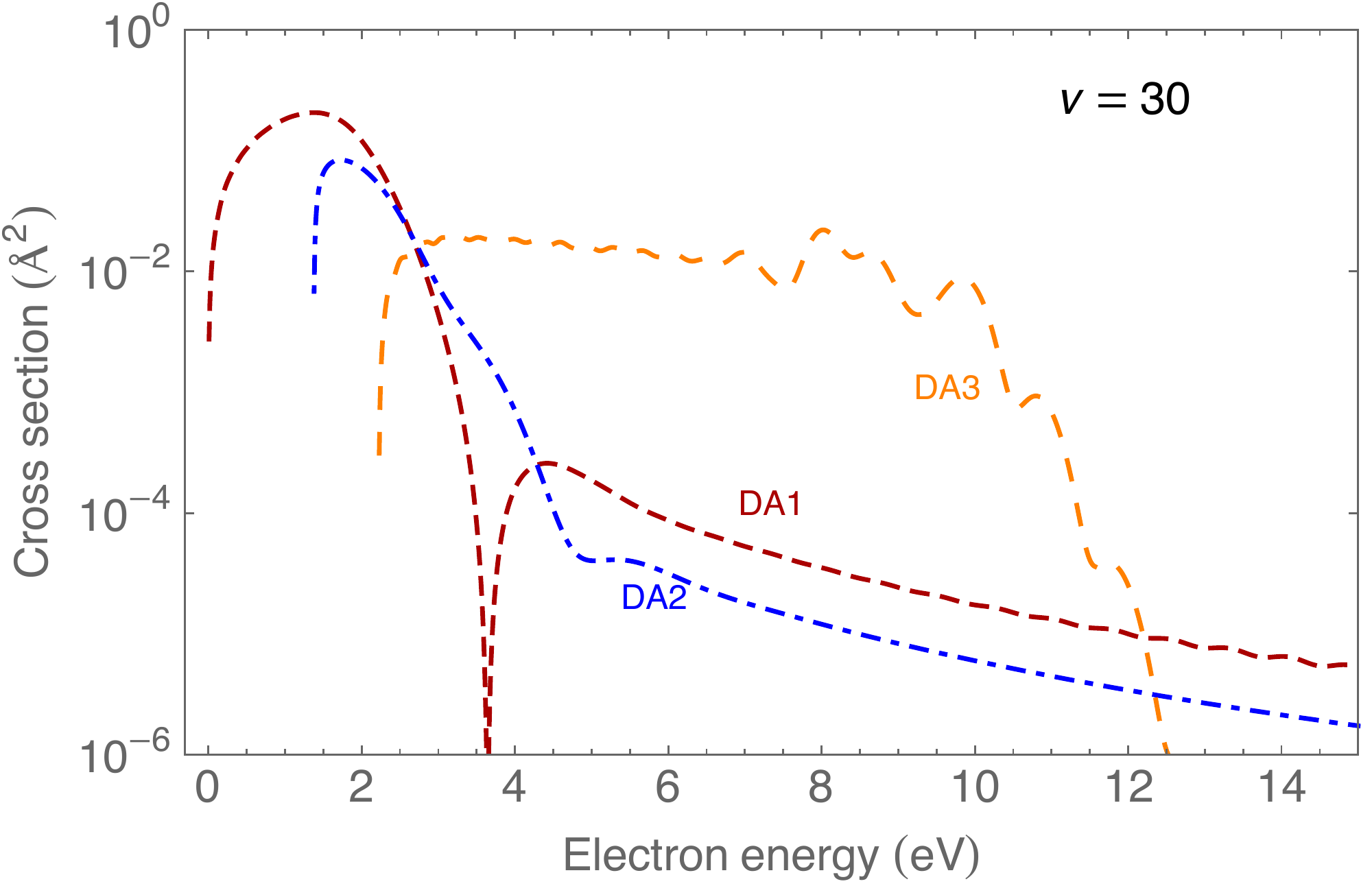} & \includegraphics[scale=.28]{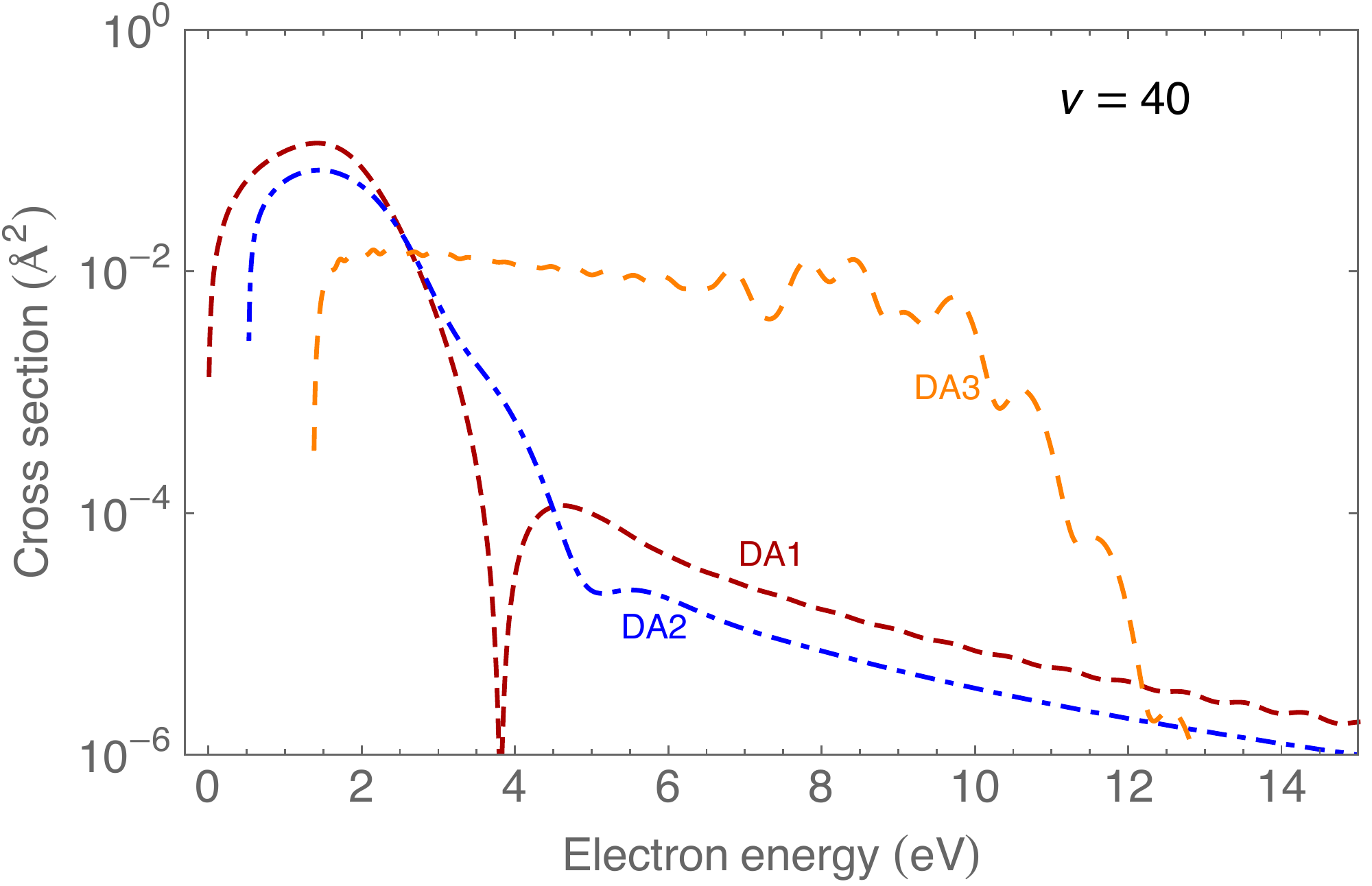} & \includegraphics[scale=.28]{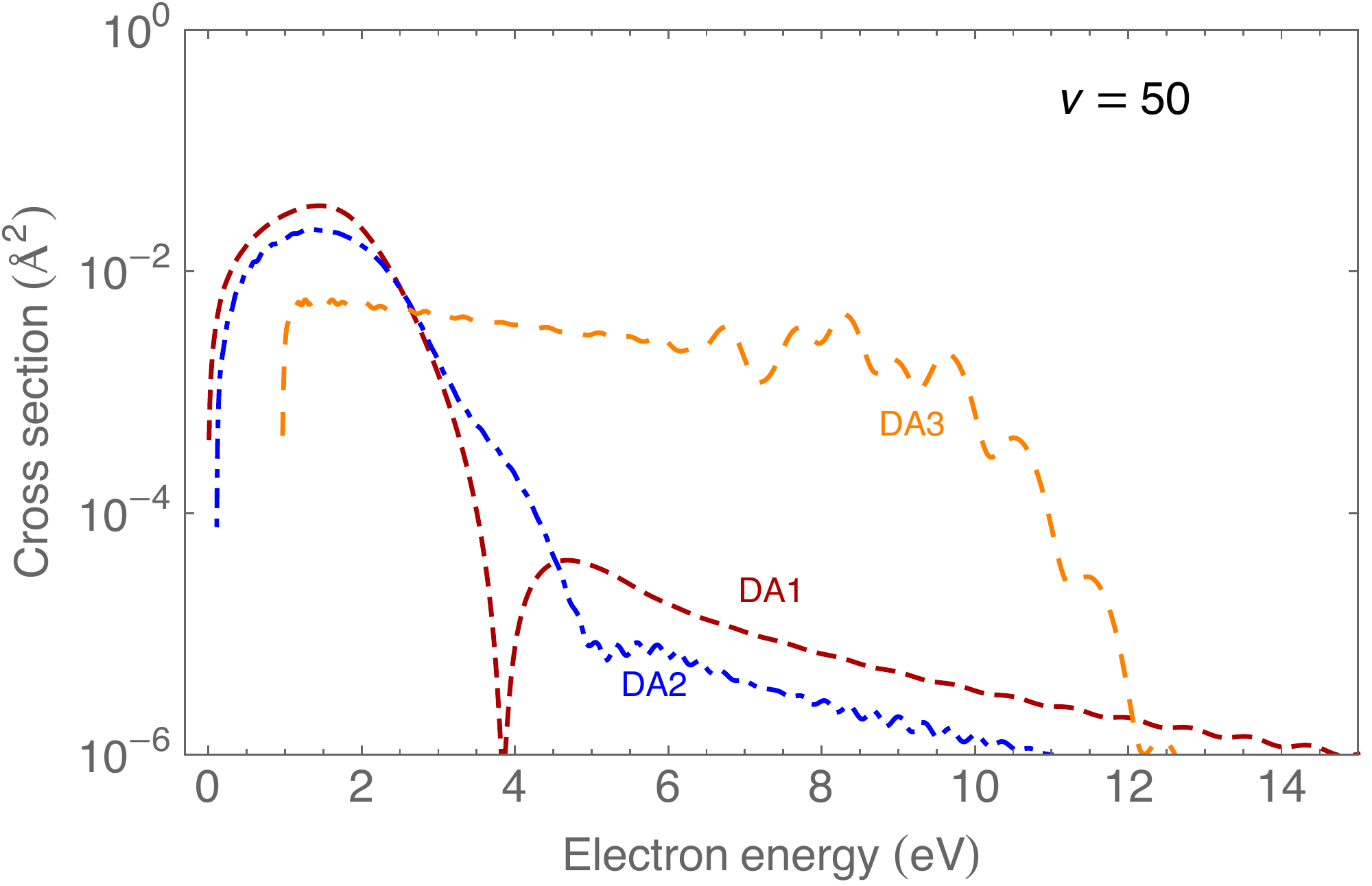} 
\end{tabular}
\caption{Overview over vibrational- and channel-resolved cross sections for NO dissociative electron attachment process \blue{for $j=0$}.  \label{fig:xsecDA_channels}}
\end{figure}
\begin{figure}
\centering
\includegraphics[scale=.4]{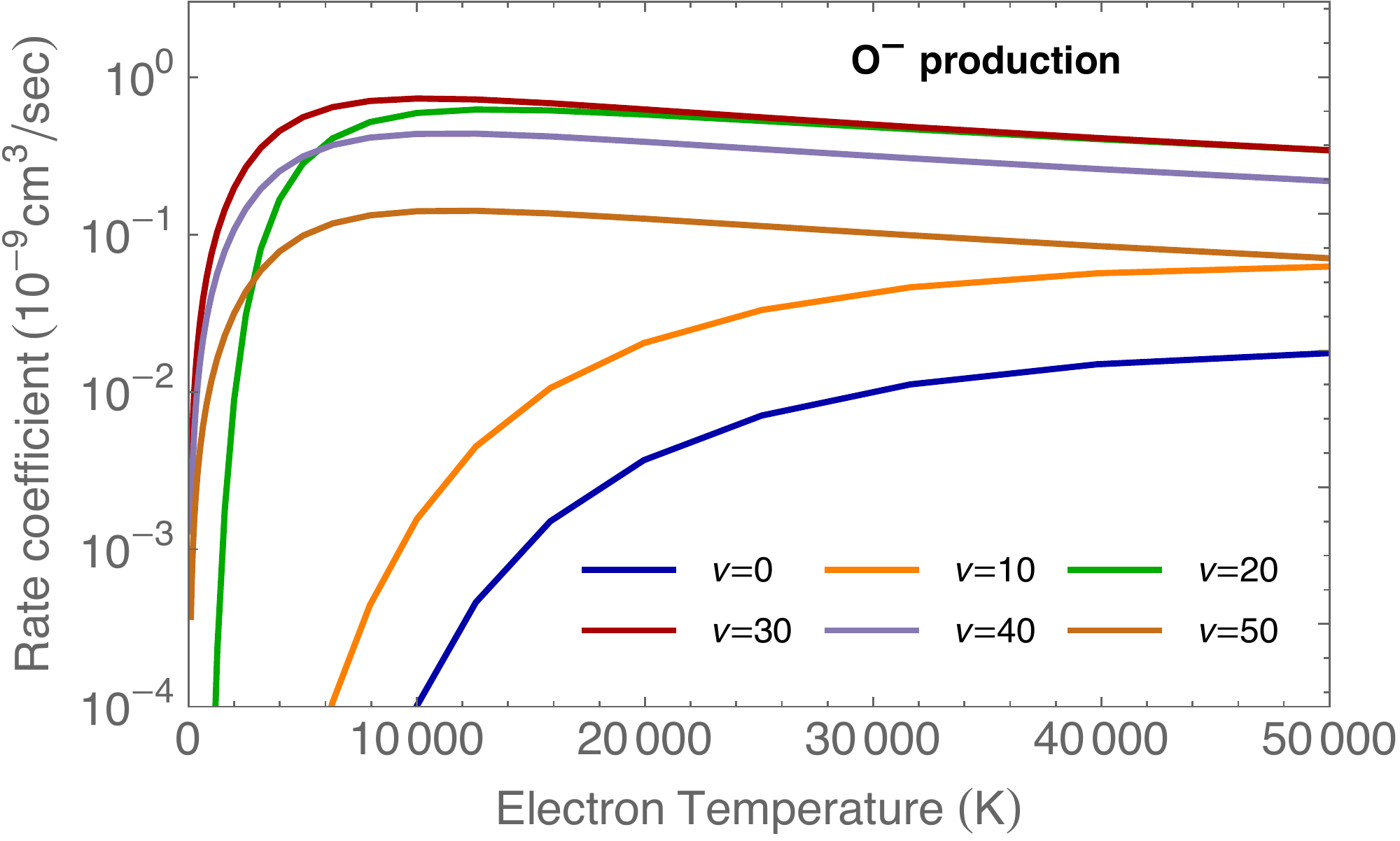} \hspace{.5cm} \includegraphics[scale=.4]{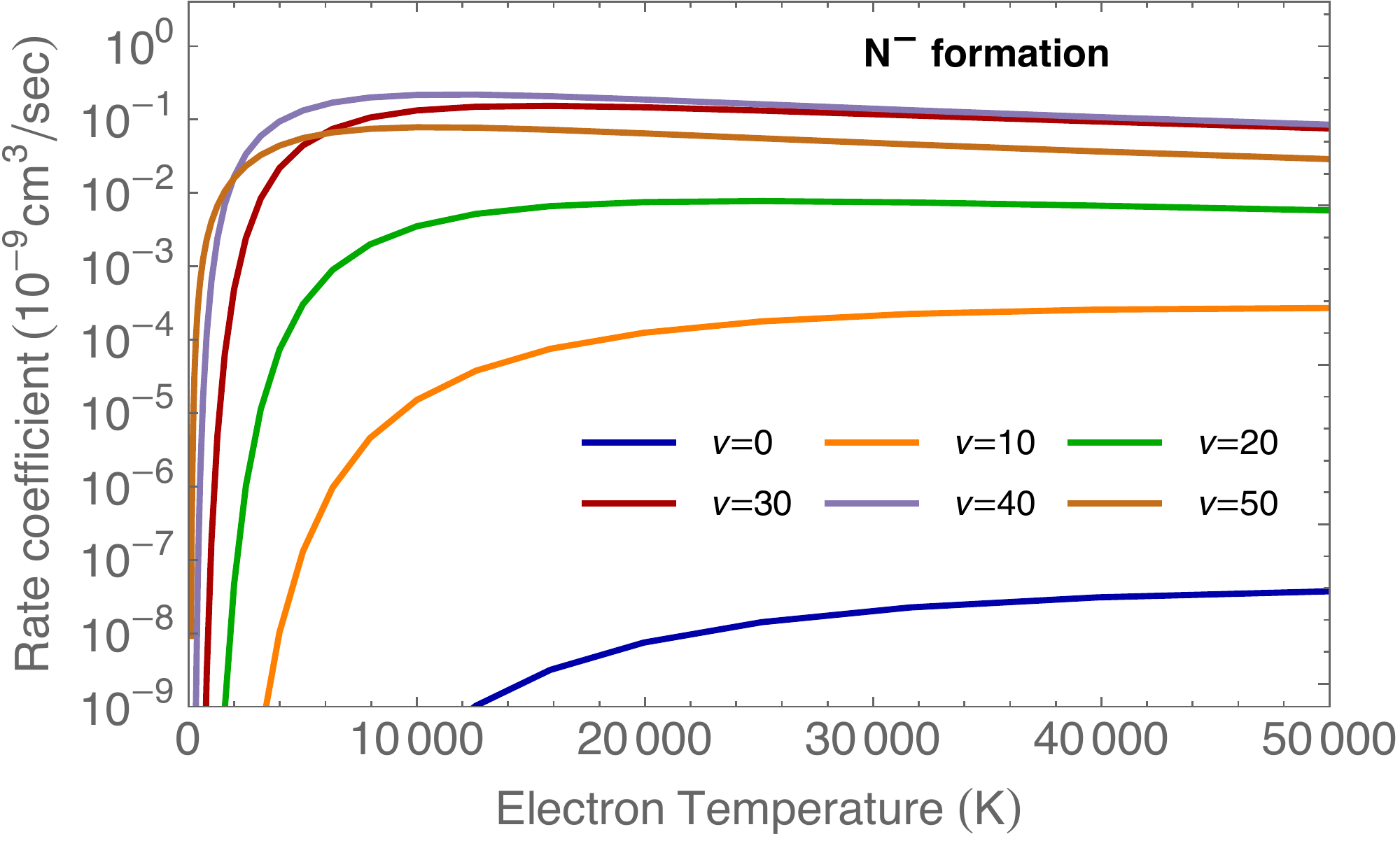} 
\caption{Summary on vibrational state-resolved rate coefficients for NO dissociative electron attachment process: on the left side, O$^-$ production sum over DA1 and DA3 channels; on the right side, N$^-$ formation coming from DA2 channel. \label{fig:rateDA_full}}
\end{figure}

Figures~\ref{fig:DAcfr_briglia}--\ref{fig:rateDA_full} contain the results of this letter. In order to validate the theoretical model, figure~\ref{fig:DAcfr_briglia} reports the comparison between the calculated DA cross section for $v=0$ with the experimental data of Rapp and Briglia ~\cite{rapp:1480}. The global agreement is quite satisfactory. Looking in detail, we note that the low energy contribution comes from the DA1 channel, which corresponds to the $\mathrm{N}(^4\mathrm{S}) + \mathrm{O}^-(^2\mathrm{P})$ fragments, both in the ground electronic state, and when the DA3 channel opens, around 7.5 eV, the cross section is dominated by the production of the excited nitrogen atom $\mathrm{N}(^2\mathrm{D})$ and of the anion $\mathrm{O}^-(^2\mathrm{P})$.  The contribution coming from the DA2 channel is negligible: \blue{Indeed,}   the electron affinity of the nitrogen atom is positive and very small - see Table~\ref{tab:NO-_limits} -  and, \blue{on the other hand,} for low vibrational levels the  production of N$^-$ is energetically forbidden. This is consistent with the interpretation of the DA cross section for NO given in ~\cite{doi:10.1063/1.459882, PhysRevLett.74.5017} and with the experimental observations reported in~\cite{C0CP01067G}.

Figure~\ref{fig:xsecDA_BR} displays the cross sections branching ratios (BR) for some initial vibrational states $v$. They are defined by:
\begin{equation}
\mathrm{BR}^i_v(\epsilon) = \frac{\sigma^i_v(\epsilon)}{\sum_{j=1}^3 \sigma^j_v(\epsilon)}\,, \hspace{1cm}i=1,2,3 \,,
\end{equation}
where $i$ stands for the  channel. In general, for all values of $v$, the DA1 and DA3 channels provide the major contributions to the reaction products. In particular, DA1 dominates the cross section at low and at high energies, whereas DA3 is important at intermediate energies. This is due to an interplay of two factors: (i) The DA3 threshold is higher than the DA1 one and, as a consequence, this \blue{latter} channel dominates at low energies; (ii) The resonance/autoionization continuum couplings within the  $^3\Pi$ and the $^1\Pi$ symmetry  are smaller than those \blue{within} the $^3\Sigma^-$ symmetry, and they are confined to a narrow energy range  around the DA3 threshold,  meaning that the contribution of the channel (\ref{eq:DAprocess3}) drops off quickly at higher energy.  The DA2 channel becomes more prominent as the NO vibrational level, $v$, increases. The importance of the DA2 channel depends on energy as does that of the DA1 one, as a consequence of the similarity  of the couplings characterizing the states to which they are correlated - see rhs of figure~\ref{fig:NOpot}. The DA2 contribution becomes evident close to the N$^-$ threshold and at high energies: In general its contribution ranges from 20~\% to 40~\% of the total cross section. In particular, for $v=20$, we notice an enhancement of the BR close to 3~eV, where the Franck-Condon factors are particularly favorable, and where the DA2 channel becomes dominant in the DA of NO and, consequently, in the production of N$^-$ anions.

Figure~\ref{fig:xsecDA_channels} provides an overview over  vibrational - and channel-resolved absolute cross sections for NO dissociative electron attachment. \blue{To validate the numerical results, we varied the coefficients in table~1 of \cite{10.1088/1361-6595/ab86d8} by $\pm10\%$ and found that the effect on the cross sections is negligible.} Figure~\ref{fig:rateDA_full} reports the full sets of results for O$^-$ production (DA1 and DA3 channels) and for N$^-$ formation (DA2 channel) -- resolved over the vibrational ladder --  for DA rate coefficient  $K_v$, by assuming a Maxwellian distribution for the electrons at temperature $T_e$:
\begin{equation}
K_v(T_e) = \left(\frac{1}{m\,\pi}\right)^{1/2}\,\left(\frac{2}{k_B T_e}\right)^{3/2}\,\int\,\epsilon\,\sigma_v(\epsilon)\,e^{-\epsilon/k_B T_e}\,d\epsilon\,.
\end{equation}
Since  three channels contribute to the full cross section, and the dissociative paths open at different thresholds, the trend of the rate coefficient as a function of the vibrational level $v$ is not regular. 

\begin{figure}
\centering
\begin{tabular}{ccc}
\includegraphics[scale=.28]{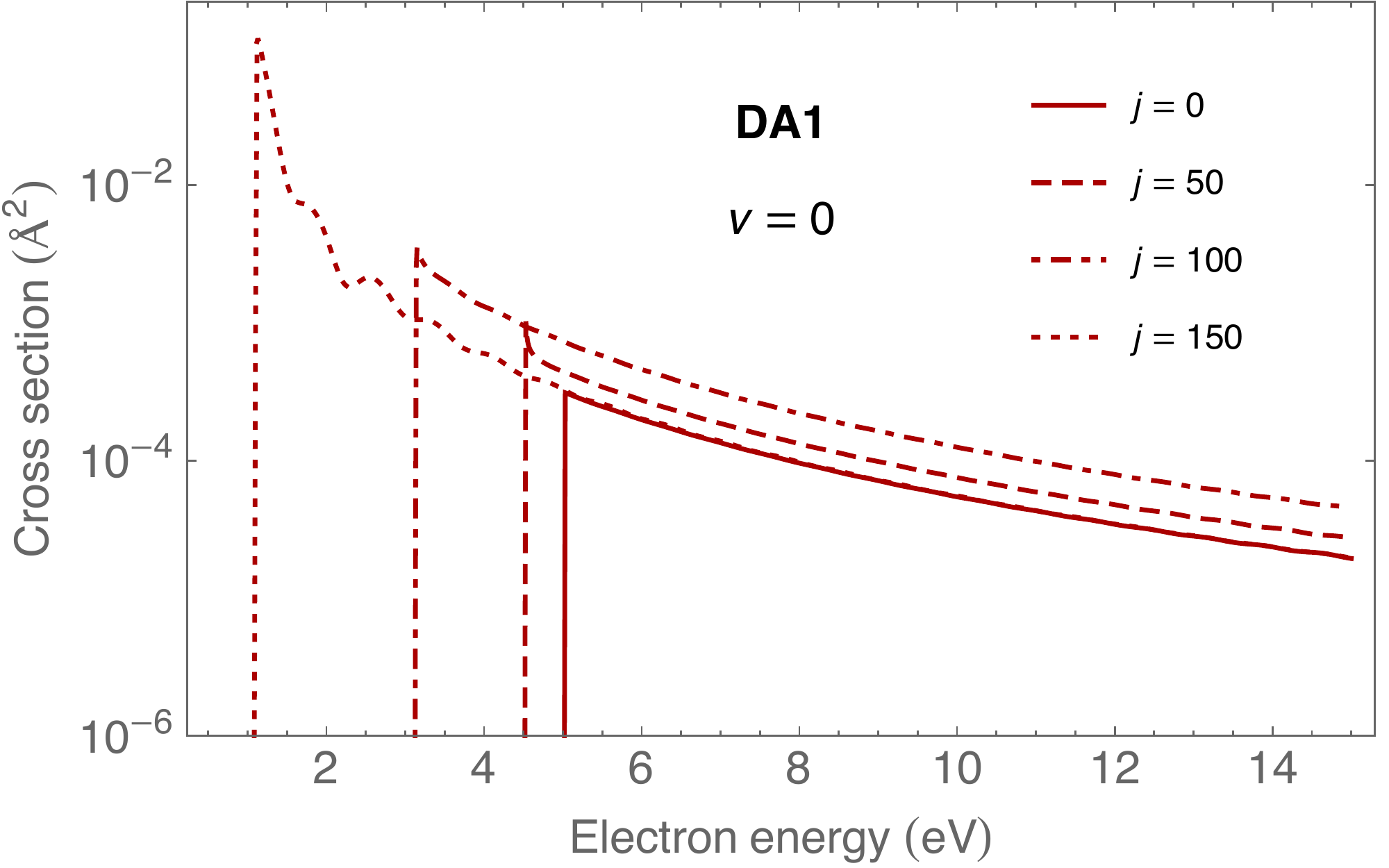} & \includegraphics[scale=.28]{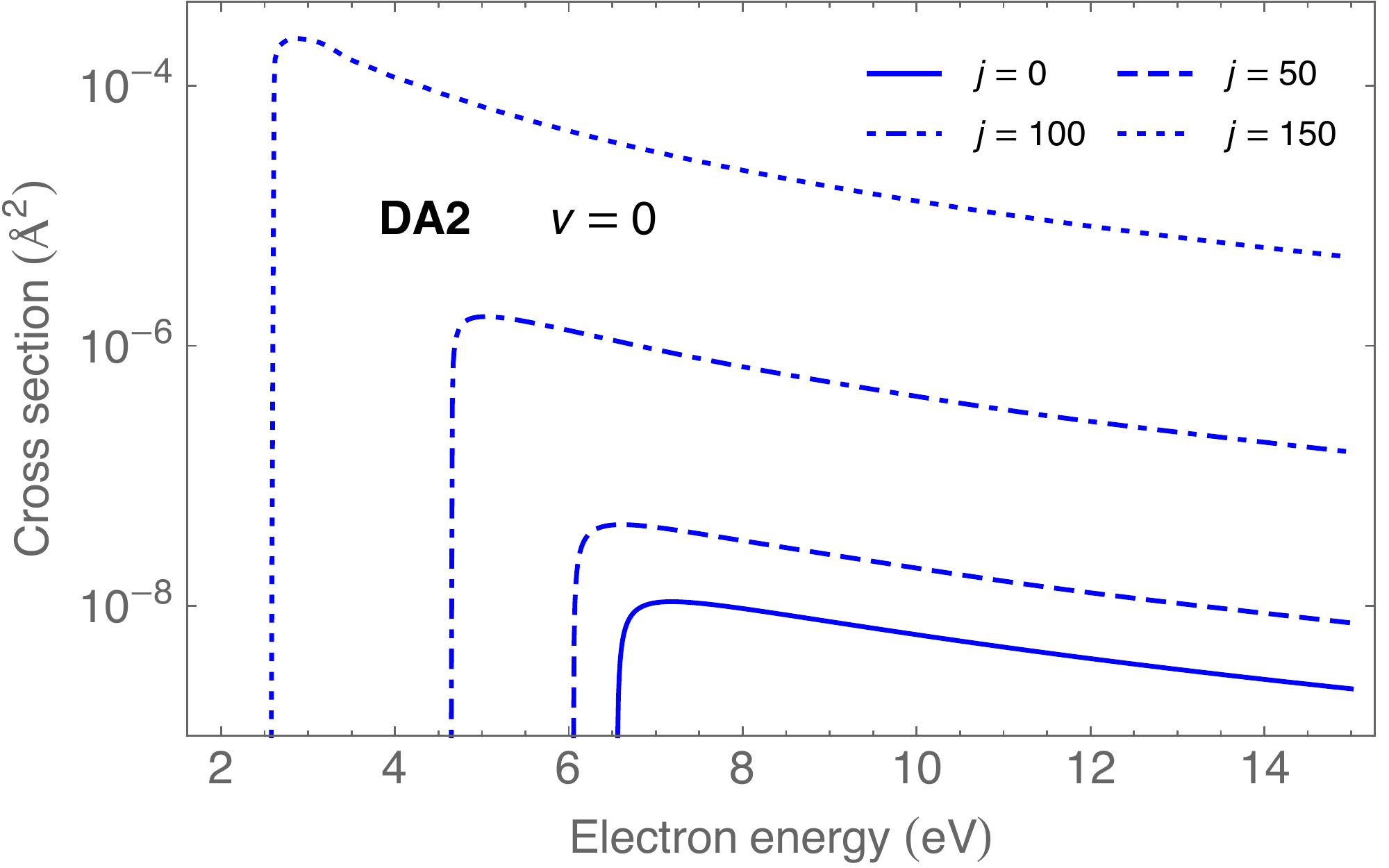} & 
\includegraphics[scale=.28]{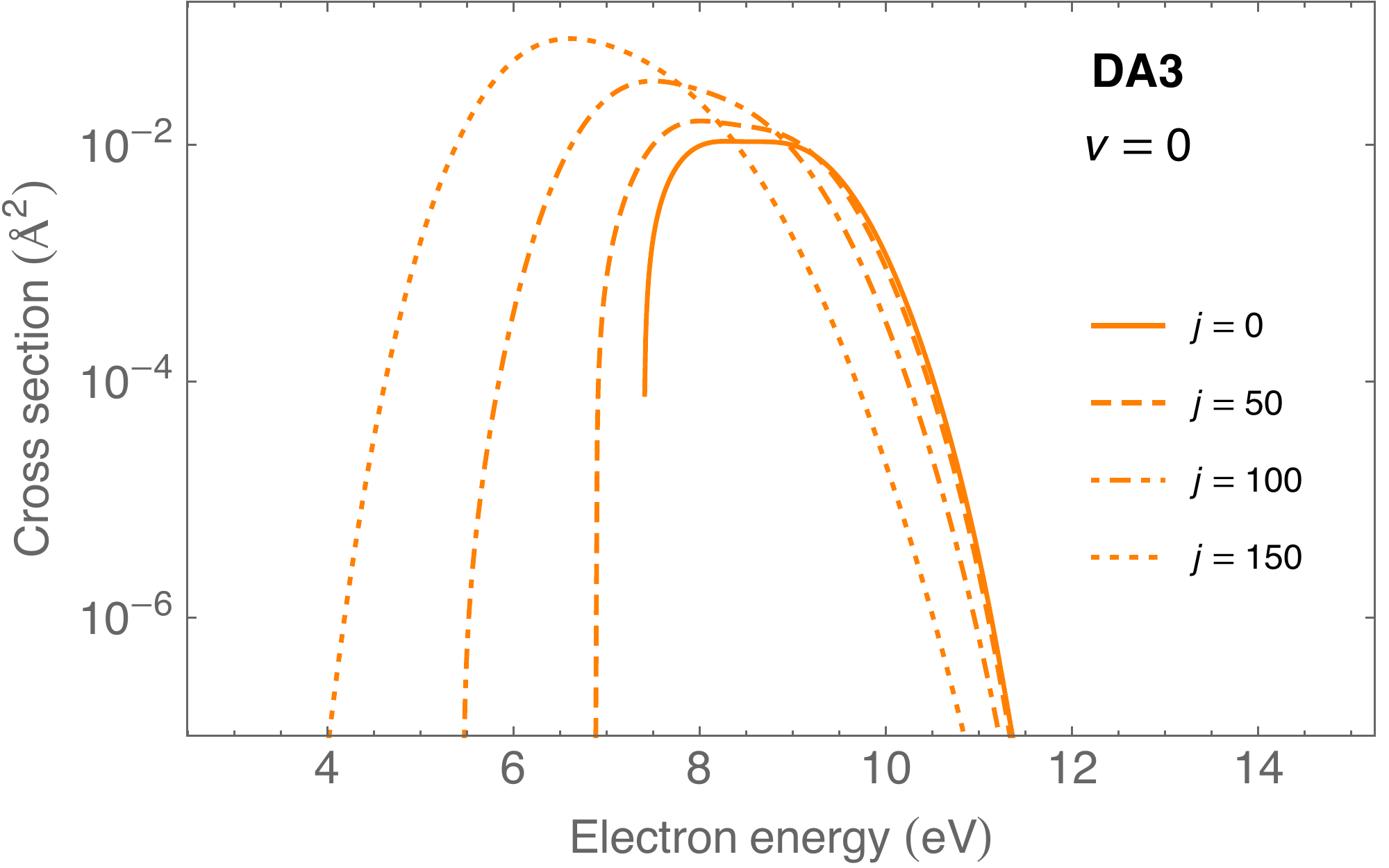}
\end{tabular}
\caption{\blue{Effects of molecular rotation on cross sections for $v=0$.} \label{fig:xsecDA_jrot}}
\end{figure}

\blue{Finally, figure~\ref{fig:xsecDA_jrot} shows the effects of molecular rotation on the cross sections for the three channels DA1, DA2 and DA3, for $v=0$. As expected, for all three processes, rotational excitation generates a threshold-shift in the curves toward smaller energies: This behaviour is, basically, due to the fact that the depth of the potential well decreases as the rotational quantum number $j$ increases and accordingly the corresponding channel open earlier. The calculated rotationally excited cross sections are essentially insensitive to $j$ at high energies for DA1 and DA3 channels, whereas  the DA2 process exhibit a large rotational effect and the whole cross section increases by several order of magnitude as a function of $j$. The reasons for this behaviour can be traced back to the shape of the effective potentials for $j>0$ compared to that for $j=0$ and in particular on the relative positions of the NO$^-$ asymptotes compared to the NO asymptote: for the DA2 process the NO$^-(^1\Delta)$, and NO$^-(^1\Sigma^+)$ thresholds are very close to the NO threshold, the effective potentials for $j>0$ are a `compressed-copy'  of the $j=0$ one and this drives an enhancement in the cross sections - \textit{i.e.} same shape but at higher energies. Conversely, rotational excitation distorts the shape of the effective potentials and this effect disappears.}


In conclusion, in this letter we presented \blue{ro-}vibrational state-resolved cross sections of the dissociative electron attachment process for the nitric oxide computed by using the theoretical formalism of the Local-Complex-Potential. We focused on three different channels of scattering (\ref{eq:DAprocess1}-\ref{eq:DAprocess3}), which result in the production of excited nitrogen atoms and of N$^-$ anions. The principal results can be outlined as follows:
\begin{itemize}
\item[(i)] The major contribution to the experimental DA cross section for $v=0$ is due to the production of nitrogen atoms in the excited state N$(^2\mathrm{D})$;
\item[(ii)] An important amount of N$^-$ ions is produced in the DA of vibrationally excited NO molecules;
\blue{\item[(iii)] Large rotational effects are found in particular for DA2 channel where the corresponding cross section increases by several orders of magnitude compared to that for $j=0$.}
\end{itemize}
The full set of data obtained in the present work and in~\cite{10.1088/1361-6595/ab86d8} are available on LXCat database (www.lxcat.net/Laporta) \cite{doi:10.1002/ppap.201600098}.

\section*{Acknowledgements}

VL and IFS acknowledge support from the French agencies ANR {\it via} the project MONA; CNRS,  CEA and CNES {\it via} the PCMI program of INSU; FR-FCM (CNRS, CEA and Eurofusion), La R\'egion Normandie, FEDER and LabEx EMC$^3$ {\it via} the projects Bioengine, EMoPlaF and CO$_2$-VIRIDIS; COMUE Normandie Universit\'e, FR-IEPE and the European Union {\it via} the COST action MD-GAS (CA18212). VL thanks the Laboratoire Ondes et Milieux Complexes (CNRS-Universit\'{e} Le Havre Normandie, France), where this work was started, for the kind hospitality.


\end{document}